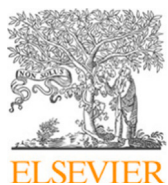

Contents lists available at ScienceDirect

# Finance Research Letters

journal homepage: www.elsevier.com/locate/frl

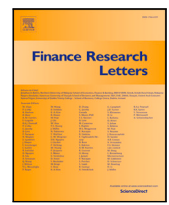

# The impact of access to credit on energy efficiency

Jun Zhou [a], Zhichao Yin [b], Pengpeng Yue [a],*

[a] *Beijing Technology and Business University, Beijing 100048, China*
[b] *Capital University of Economics and Business, Beijing, 100070, China*

ARTICLE INFO

ABSTRACT

This paper proposes a brand-new measure of energy efficiency at household level and explores how it is affected by access to credit. We calculate the energy and carbon intensity of the related sectors, which experience a substantial decline from 2005 to 2019. Although there is still high inequality in energy use and carbon emissions among Chinese households, the energy efficiency appears to be improved in long run. Our research further maps the relationship between financial market and energy. The results suggest that broadened access to credit encourages households to improve energy efficiency, with higher energy use and carbon emission.

## 1. Introduction

A large strand of literature has focused on household energy use and related carbon emissions. Many researchers posit that household carbon emissions account for a significant proportion of the national total emissions (Wei et al., 2007; Wang and Yang, 2014; Li et al., 2019a). Wei et al. (2007) posit that approximately 26 percent of total energy consumption and 30 percent of $CO_2$ emission every year are due to household lifestyles. Compared to the direct energy use and carbon emissions, indirect parts are much more salient (Wang and Yang, 2014). Following Bin and Dowlatabadi (2005) and Wei et al. (2007), we use the method of a Consumer Lifestyle Approach (CLA) to estimate the energy intensity and carbon intensity of consumer expenditure in all relevant sectors during 2005 and 2019, based on which the proportions that household consumption contributes to energy use and carbon emissions are clarified. This paper aims to contribute to this literature by measuring energy efficiency from a new view of lower carbon emissions when consuming a given unit of energy and elicit its trends. Furthermore, recent studies indicate a potential impact of financial market on the volume or the efficiency of energy use from the firm to the regional level (Le et al., 2020; Zhang et al., 2020; Yu et al., 2022). Consistently, we try to examine the role access to credit plays in household energy efficiency. Our goal is to map the relationships between financial market and energy in the light of household evidence, which is largely unexplored.

Prior literature on household energy consumption mostly centers on energy use or carbon emission (Wei et al., 2007; Wu et al., 2017; Li et al., 2019a). Energy efficiency attracts increasing attention in response to energy crisis and climate change (Linares and Labandeira, 2010). Consumers are often considered ignorant of the extent to which energy implications of their purchases can contribute to energy efficiency (Boardman, 2004); nevertheless, their increased demand causes rapid expansion in energy requirements (Dubois et al., 2019; Li et al., 2019b, 2021). In fact, the household sector is energy-intensive and its energy efficiency gives cause for concern (Solà et al., 2021). Referring to the better way of using energy, energy efficiency involves diversified indicators (Herring, 2006). The regional energy efficiencies are evaluated under different technology assumptions (Zhang et al., 2013; Wang and Wei, 2014). On the supply side, it has traditionally been defined as the technical ratio between energy input and output (Oikonomou et al., 2009). Energy efficiency is also discussed in the realms of literature on market barriers (Reddy, 1991), energy policies (Gillingham et al., 2006), energy conservation (Oikonomou et al., 2009), energy poverty (Nguyen and Nasir, 2021) and energy efficiency gap (Jaffe and Stavins, 1994).

Despite the strands of literature about better way of using energy, Herring (2006) argues that improving energy efficiency lowers the implicit price of energy, which finally leads to greater use. With more energy use comes more carbon emissions. Although carbon emissions and energy use both matter in household lifestyles (Le et al., 2020; Wu et al., 2017; Li et al., 2019a), the traditional indicators of energy efficiency seem to be of limited use because of ignoring carbon emissions. And there are fewer studies focusing on energy efficiency in the perspective of the energy intensity and carbon intensity. Following Patterson (1996), Linares and Labandeira (2010), and Wei et al. (2007), based on carbon and energy intensity, we construct a new indicator of energy efficiency, the ratio of carbon emissions to energy use of household, which could be used in exploring ways to reduce carbon emissions.

Recent research has noticed the role of financial development in energy issues. Le et al. (2020) posit that financial inclusion leads to higher emissions of CO2 in the region. Zhang et al. (2020) find firms with access to credit are associated with lower energy efficiency of Chinese manufacturing firms, while Yu et al. (2022) assert digital finance, along with its coverage breadth and usage depth, significantly improved renewable energy consumption in China. Motivated by previous findings, this study investigates the impact of access to credit on energy use, carbon emissions, and the energy efficiency due to household lifestyles.

Following Wei et al. (2007), we calculate the energy intensity and carbon intensity of consumer expenditure in all relevant sectors using China Energy Statistical Yearbook (CESY, 2006–2020), and we find the carbon and energy intensity show a descending trend from 2005 to 2019. There is a wide disparity in energy use and carbon emissions among Chinese households though. Using a biennial micro dataset from China Household Finance Survey (CHFS), the empirical results show that household access to credit will lead to a higher energy use and carbon emissions simultaneously, but encourage households to improve energy efficiency eventually. And we also find there is a non-linear relationship between energy efficiency and access to credit when taking the square of explanatory variable into account.

This paper contributes to the literature by introducing a novel method to measure energy efficiency, which reflects lower carbon emissions generated by energy use from household consumption. Existing research considers energy efficiency as the technical ratio between energy input and output (Oikonomou et al., 2009). Scant attention has been paid to energy efficiency from perspectives of carbon emissions by household consumption behaviors. To measure energy efficiency at the household level, we provide the energy intensity and the carbon intensity of the related sectors in 2005–2019, which are available to be used directly to estimate energy use and carbon emissions. Hence, the new measurement has wide applicability to describe the long-run trends of energy efficiency in extensive contexts. And recently, another strand of literature focuses on how financial development affects energy consumption in an industry or in a region (Le et al., 2020; Zhang et al., 2020). This paper also fills in the important gap for the link between financial market and energy by identifying the impact of financial inclusion on household energy efficiency, which aims to guide policymakers in facilitating the financial markets as tools to meet carbon emission goals.

## 2. Research methodology

### 2.1. Data and variables

This study exploits data from CESY (2006–2020), China Industrial Statistical Yearbook (CISY, 2006–2020), China Statistical Yearbook (CSY, 2006–2020), China Population Statistical Yearbook (CPSY, 2006–2020), and CHFS (2011, 2013, 2015, 2017, 2019). Following Wei et al. (2007), we adopt the method of a Consumer Lifestyle Approach (CLA) to calculate the energy intensity (See Table A.1) and the carbon intensity (See Table A.2) of consumer expenditure, and then estimate the indirect energy use and carbon emissions by household consumption behaviors in 2005–2019. In this paper, energy use and carbon emissions are generated in the production process of products and services to meet demand for household consumption in food, clothing, facilities, medicine, communication, education, residence and commodities. Tables A.1 and A.2 show the energy and carbon intensity, which vary across the related sectors of consumer expenditure, falling significantly between 2005 and 2019.

To measure the indirect energy use and carbon emissions at household level, we use the consumption of related sectors to multiply the energy intensity and the carbon intensity of related sectors, respectively. Using CHFS, we aggregate the household energy use and carbon emissions for all consumption categories. Then, we construct a new indicator of energy efficiency, the ratio between carbon emissions and energy use of household, to reflect carbon emissions upon consuming per unit of energy. For the independent variable of interest, we use the total amount of loans that household is still in debt as the proxy of access to credit. With reference to prior literature, control variables used at household level include age, male/female classification, years of schooling, married, employed, health, income, wealth, business, family size, and rural/urban residence classification. Table 1 reports summary statistics for variables used.

**Table 1**
Summary statistics.

| | Obs | Mean | Std. | Min. | Max. |
|---|---|---|---|---|---|
| Energy Use | 133,122 | 72.365 | 148.497 | 0.043 | 7588.833 |
| Carbon Emissions | 133,122 | 654.155 | 1334.875 | 0.382 | 68,856.930 |
| Energy Efficiency | 133,122 | 9.046 | 0.154 | 8.618 | 9.371 |
| Credit Access | 133,122 | 34 709.629 | 120,276.304 | 0.000 | 1,000,000 |
| Age | 133,122 | 52.807 | 13.465 | 16 | 80 |
| Male | 133,122 | 0.768 | 0.422 | 0 | 1 |
| Schooling | 133,122 | 9.479 | 4.205 | 0 | 22 |
| Married | 133,122 | 0.864 | 0.343 | 0 | 1 |
| Employed | 133,122 | 0.671 | 0.470 | 0 | 1 |
| Health | 133,122 | 0.418 | 0.493 | 0 | 1 |
| Income | 133,122 | 77,178.885 | 94,990.140 | 1.050 | 690,180 |
| Wealth | 133,122 | 903,076.875 | 1,503,695.659 | 4.000 | 979,4016 |
| Business | 133,122 | 0.141 | 0.348 | 0 | 1 |
| Family Size | 133,122 | 3.332 | 1.592 | 1 | 20 |
| Rural | 133,122 | 0.325 | 0.469 | 0 | 1 |

This table presents summary statistics. Energy Use is the indirect energy use of household. Carbon Emissions is the carbon emissions of household. Energy Efficiency is the carbon emissions per unit of energy for a household. A variety of variables at household level are controlled. For each variable, this table reports the number of observations, mean, standard deviation, minimum, and maximum values.

*2.2. Models*

We compile a panel dataset using CHFS in the regression tests. This paper examines the impact of access to credit on energy use of household using Eq. (1):

$$ln(\text{Energy Use})_{it} = \alpha + \beta ln(\text{Credit Access})_{it} + X_{it}\gamma + c_i + \mu_{it} \quad (1)$$

where Energy Use$_{it}$ is the indirect energy use of household, and Credit Access$_{it}$ represents the amount of loans for household *i*.

This paper examines the impact of access to credit on carbon emissions of household using Eq. (2):

$$ln(\text{Carbon Emissions})_{it} = \alpha + \beta ln(\text{Credit Access})_{it} + X_{it}\gamma + c_i + \mu_{it} \quad (2)$$

where Carbon Emissions$_{it}$ is the carbon emissions of household.

This paper further investigates the impact of access to credit on energy efficiency of household using Eq. (3):

$$ln(\text{Energy Efficiency})_{it} = \alpha + \beta ln(\text{Credit Access})_{it} + X_{it}\gamma + c_i + \mu_{it} \quad (3)$$

where Energy Efficiency$_{it}$ is the carbon emissions per unit of energy for a household, the lower value corresponding to higher energy efficiency.

## 3. Inequality among Chinese household and dynamics of Energy Efficiency

Inequality has become an important topic in the field of energy economics (Nguyen and Nasir, 2021). Based on Gini coefficients, Fig. 1 shows the inequality of income, consumption, energy use and carbon emissions among Chinese households in 2019. The richest 10% of population owns 37.97% of total household income, 22.29% of total household consumption, 23.85% of total household energy use, and 23.88% of total household carbon emissions. In comparison, the poorest 10% of residents only account for 0.38% of total household income, 5.09% of total household consumption, 4.90% of total household energy use, and 4.89% of total household carbon emissions. It means there is high inequality in energy use and carbon emissions among Chinese households.

Fig. 2 presents the dynamics of household energy efficiency. As mentioned above, energy efficiency reflects carbon emissions upon consuming per unit of energy, where lower value corresponds to greater efficiency. It shows that the energy efficiency appears to be improved from 2005 to 2019. And the households in rural areas have a better level of energy efficiency than those in urban areas. It makes sense because the consumption basket of households in urban areas consists of more expenditure in energy-intensive categories such as educations and residence. The significant amelioration of energy efficiency holds altogether.

## 4. Access to credit improves Energy Efficiency of household

We examine the impact of access to credit on household energy use, carbon emissions, and energy efficiency by using Eq. (1) in Columns (1) and (2), Eq. (2) in Columns (3) and (4), and Eq. (3) in Columns (5) and (6) of Table 2. The fixed-effect (FE) estimates indicate that a 1% increase in the household loan will lead to a 1.192% increase in household energy use, a 1.188% increase in household carbon emissions, and a 0.004% decrease of carbon emissions per unit of energy. In unreported tests, we use repeated households sample for robustness checks. The results consistently suggest that access to credit market will encourage households to improve energy efficiency, when both the energy use and carbon emissions of household increase.

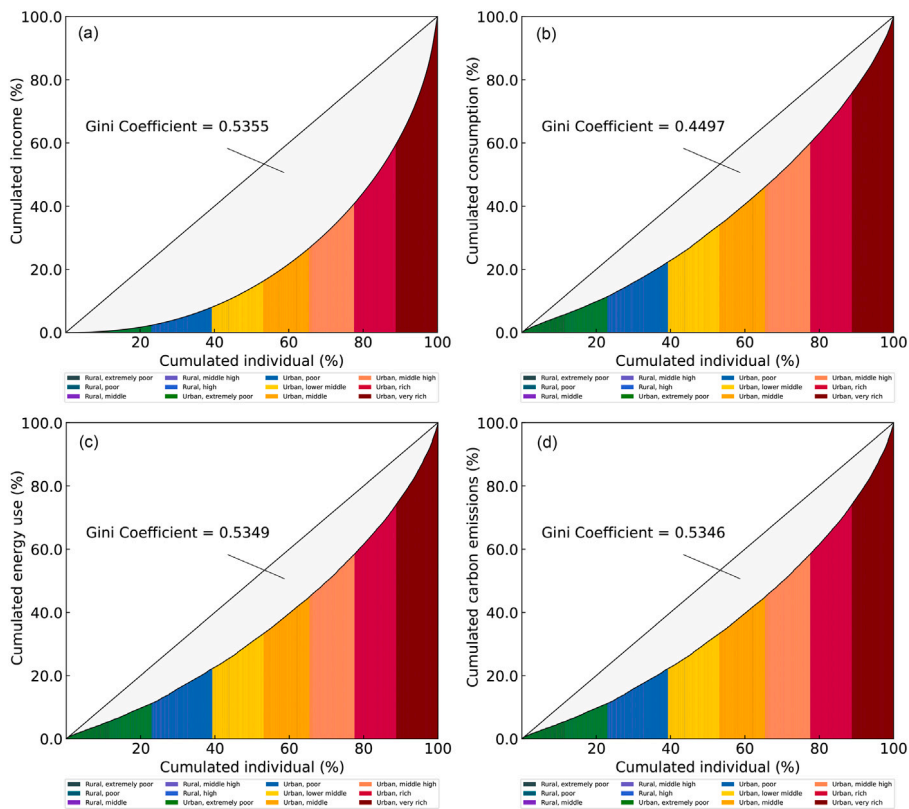

**Fig. 1.** Inequality among Chinese households. (a) Lorenz curve of household income. (b) Lorenz curve of household consumption. (c) Lorenz curve of household energy use. (d) Lorenz curve of household carbon emissions.

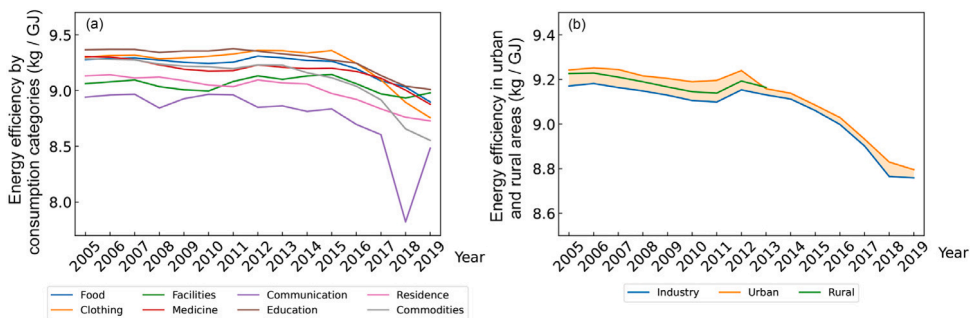

**Fig. 2.** The dynamics of energy efficiency of household in 2005–2019. (a) Energy efficiency by consumption categories. (b) Energy efficiency in urban and rural areas.

It has been documented that access to credit matters in household consumption (Yue et al., 2022). Easier access to credit could provide consumers with extended options in changing lifestyles which further determine the energy use and the related carbon emissions (Wei et al., 2007). Since energy and carbon intensity of consumer expenditure vary across different sectors, the energy efficiency may be altered when households try to abstain from energy/carbon-intensive behaviors and adjust their purchasing choices in different categories of consumer goods. The patterns households use energy can be spontaneously optimized in a more efficient way, which implies households energy efficiency gets improved as a consequence of broadening access to credit.

Table 3 shows there is a non-linear relationship between energy use, carbon emissions, energy efficiency and household access to credit when introducing the square of access to credit in regressions. Following Jappelli (1990), we plot the energy use, carbon emissions, and energy efficiency against household debts when all variables are kept constant at their sample mean values in Fig. 3. The energy use and carbon emissions are both on the rise, while the emissions per unit of energy is on the decline, which declines by more than 0.03% (0.003/9.162) from low levels of household loans (1000 RMB) to values above 100,000 RMB. It means energy efficiency is improving, with energy use and carbon emissions climbing.

**Table 2**
Access to credit helps to improve energy efficiency.

|  | $ln$(Energy Use) | | $ln$(Carbon Emissions) | | $ln$(Energy Efficiency) | |
| --- | --- | --- | --- | --- | --- | --- |
|  | (1) OLS | (2) FE | (3) OLS | (4) FE | (5) OLS | (6) FE |
| $ln$(Credit Access) | 0.01272*** | 0.01192*** | 0.01268*** | 0.01188*** | −0.00004*** | −0.00004*** |
|  | (0.00048) | (0.00078) | (0.00048) | (0.00078) | (0.00000) | (0.00001) |
| Age | −0.00434*** | −0.01018*** | −0.00442*** | −0.01027*** | −0.00008*** | −0.00009*** |
|  | (0.00110) | (0.00318) | (0.00110) | (0.00318) | (0.00001) | (0.00003) |
| $Age^2$/100 | −0.00496*** | 0.00371 | −0.00492*** | 0.00378 | 0.00004*** | 0.00007*** |
|  | (0.00106) | (0.00308) | (0.00106) | (0.00307) | (0.00001) | (0.00003) |
| Male | −0.04769*** |  | −0.04798*** |  | −0.00029*** |  |
|  | (0.00507) |  | (0.00507) |  | (0.00004) |  |
| Schooling | 0.01876*** | 0.01006** | 0.01868*** | 0.01003** | −0.00008*** | −0.00003 |
|  | (0.00174) | (0.00390) | (0.00174) | (0.00390) | (0.00001) | (0.00003) |
| $Schooling^2$/100 | 0.04063*** | −0.01503 | 0.04159*** | −0.01478 | 0.00096*** | 0.00026 |
|  | (0.00868) | (0.02159) | (0.00868) | (0.02156) | (0.00007) | (0.00019) |
| Married | 0.12868*** | 0.05972*** | 0.12869*** | 0.05955*** | 0.00001 | −0.00017 |
|  | (0.00682) | (0.01504) | (0.00682) | (0.01502) | (0.00006) | (0.00013) |
| Employed | −0.16293*** | −0.03595*** | −0.16310*** | −0.03607*** | −0.00018*** | −0.00013* |
|  | (0.00549) | (0.00903) | (0.00548) | (0.00901) | (0.00005) | (0.00008) |
| Health | −0.00391 | −0.03372*** | −0.00396 | −0.03367*** | −0.00005 | 0.00006 |
|  | (0.00433) | (0.00674) | (0.00432) | (0.00673) | (0.00004) | (0.00006) |
| $ln$(Income) | −0.26614*** | −0.14456*** | −0.26653*** | −0.14455*** | −0.00039*** | 0.00001 |
|  | (0.01117) | (0.01590) | (0.01117) | (0.01589) | (0.00009) | (0.00014) |
| $ln$(Income)$^2$ | 0.02121*** | 0.01194*** | 0.02124*** | 0.01194*** | 0.00003*** | 0.00001 |
|  | (0.00060) | (0.00087) | (0.00060) | (0.00087) | (0.00000) | (0.00001) |
| $ln$(Wealth) | −0.15019*** | −0.05630*** | −0.15075*** | −0.05638*** | −0.00055*** | −0.00008 |
|  | (0.01253) | (0.02097) | (0.01252) | (0.02095) | (0.00010) | (0.00018) |
| $ln$(Wealth)$^2$ | 0.01210*** | 0.00604*** | 0.01212*** | 0.00604*** | 0.00002*** | −0.00000 |
|  | (0.00053) | (0.00091) | (0.00053) | (0.00091) | (0.00000) | (0.00001) |
| Business | 0.13026*** | 0.07778*** | 0.12967*** | 0.07755*** | −0.00058*** | −0.00023** |
|  | (0.00632) | (0.01224) | (0.00631) | (0.01221) | (0.00006) | (0.00011) |
| Family Size | 0.08407*** | 0.08879*** | 0.08451*** | 0.08928*** | 0.00044*** | 0.00050*** |
|  | (0.00160) | (0.00346) | (0.00160) | (0.00345) | (0.00001) | (0.00003) |
| Rural | −0.27362*** | −0.13296*** | −0.27375*** | −0.13235*** | −0.00013*** | 0.00061** |
|  | (0.00555) | (0.02938) | (0.00554) | (0.02933) | (0.00005) | (0.00026) |
| Province | Yes |  | Yes |  | Yes |  |
| Year | Yes | Yes | Yes | Yes | Yes | Yes |
| Observations | 133,122 | 133,122 | 133,122 | 133,122 | 133,122 | 133,122 |
| Adjusted $R^2$ | 0.4551 | 0.1109 | 0.4576 | 0.1148 | 0.8642 | 0.8628 |

This table presents the results of our tests analyzing the impact of access to credit on energy use, carbon emissions and energy efficiency of household.
*Indicates significance at the 10% level.
**Indicates significance at the 5% level.
***Indicates significance at the 1% level.

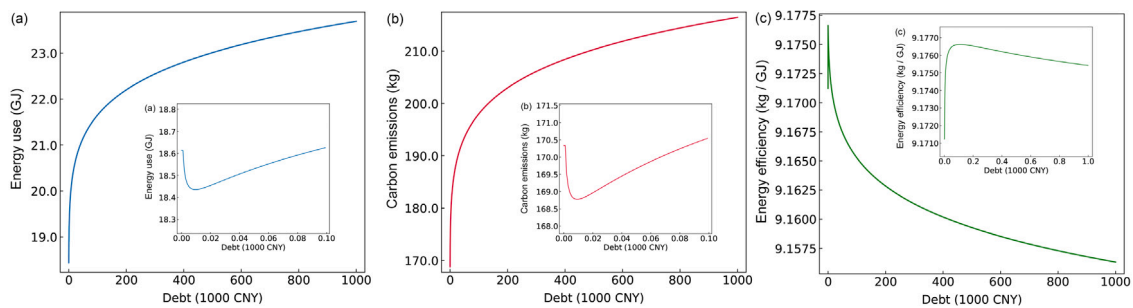

**Fig. 3.** The estimated energy use, carbon emissions, and energy efficiency for different values of household debts. (a) The estimated energy use. (b) The estimated carbon emissions. (c) The estimated energy efficiency (Note: The downward sloping curve corresponds to the rising energy efficiency).

## 5. Conclusion

This paper measures energy efficiency in a new way and investigates the impact of access to credit market on household energy efficiency. Our main findings shed new light on the nexus between financial market and energy, indicating that broadened access to credit encourages households to improve energy efficiency, with higher energy use and carbon emission.

**Table 3**
The non-linear relationship between access to credit and energy efficiency.

|  | $ln$(Energy Use) | | $ln$(Carbon Emissions) | | $ln$(Energy Efficiency) | |
| --- | --- | --- | --- | --- | --- | --- |
|  | (1) OLS | (2) FE | (3) OLS | (4) FE | (5) OLS | (6) FE |
| $ln$(Credit Access) | −0.00370 | −0.00850** | −0.00346 | −0.00825** | 0.00024*** | 0.00025*** |
|  | (0.00275) | (0.00413) | (0.00275) | (0.00413) | (0.00002) | (0.00004) |
| $ln$(Credit Access)$^2$ | 0.00147*** | 0.00188*** | 0.00145*** | 0.00185*** | −0.00002*** | −0.00003*** |
|  | (0.00025) | (0.00038) | (0.00025) | (0.00038) | (0.00000) | (0.00000) |
| Age | −0.00412*** | −0.01074*** | −0.00421*** | −0.01082*** | −0.00009*** | −0.00008*** |
|  | (0.00110) | (0.00319) | (0.00110) | (0.00318) | (0.00001) | (0.00003) |
| $Age^2/100$ | −0.00517*** | 0.00424 | −0.00513*** | 0.00430 | 0.00004*** | 0.00006** |
|  | (0.00106) | (0.00308) | (0.00106) | (0.00308) | (0.00001) | (0.00003) |
| Male | −0.04752*** |  | −0.04781*** |  | −0.00029*** |  |
|  | (0.00508) |  | (0.00507) |  | (0.00004) |  |
| Schooling | 0.01888*** | 0.01017*** | 0.01880*** | 0.01014*** | −0.00008*** | −0.00003 |
|  | (0.00174) | (0.00390) | (0.00174) | (0.00390) | (0.00001) | (0.00003) |
| $Schooling^2/100$ | 0.03948*** | −0.01592 | 0.04045*** | −0.01565 | 0.00098*** | 0.00027 |
|  | (0.00869) | (0.02159) | (0.00868) | (0.02156) | (0.00007) | (0.00019) |
| Married | 0.12770*** | 0.05889*** | 0.12773*** | 0.05873*** | 0.00003 | −0.00016 |
|  | (0.00682) | (0.01504) | (0.00682) | (0.01502) | (0.00006) | (0.00013) |
| Employed | −0.16251*** | −0.03570*** | −0.16269*** | −0.03583*** | −0.00019*** | −0.00013* |
|  | (0.00549) | (0.00902) | (0.00548) | (0.00901) | (0.00005) | (0.00008) |
| Health | −0.00478 | −0.03454*** | −0.00482 | −0.03447*** | −0.00004 | 0.00007 |
|  | (0.00433) | (0.00674) | (0.00432) | (0.00673) | (0.00004) | (0.00006) |
| $ln$(Income) | −0.26194*** | −0.14107*** | −0.26240*** | −0.14111*** | −0.00046*** | −0.00004 |
|  | (0.01117) | (0.01587) | (0.01118) | (0.01586) | (0.00009) | (0.00014) |
| $ln$(Income)$^2$ | 0.02095*** | 0.01172*** | 0.02099*** | 0.01173*** | 0.00004*** | 0.00001 |
|  | (0.00060) | (0.00087) | (0.00060) | (0.00087) | (0.00000) | (0.00001) |
| $ln$(Wealth) | −0.14719*** | −0.05389** | −0.14779*** | −0.05400** | −0.00061*** | −0.00011 |
|  | (0.01253) | (0.02097) | (0.01252) | (0.02096) | (0.00010) | (0.00018) |
| $ln$(Wealth)$^2$ | 0.01194*** | 0.00590*** | 0.01196*** | 0.00590*** | 0.00002*** | −0.00000 |
|  | (0.00053) | (0.00091) | (0.00053) | (0.00091) | (0.00000) | (0.00001) |
| Business | 0.12824*** | 0.07600*** | 0.12769*** | 0.07580*** | −0.00055*** | −0.00020* |
|  | (0.00633) | (0.01224) | (0.00632) | (0.01221) | (0.00006) | (0.00011) |
| Family Size | 0.08431*** | 0.08871*** | 0.08474*** | 0.08921*** | 0.00043*** | 0.00050*** |
|  | (0.00160) | (0.00346) | (0.00160) | (0.00345) | (0.00001) | (0.00003) |
| Rural | −0.27220*** | −0.13288*** | −0.27236*** | −0.13227*** | −0.00016*** | 0.00061** |
|  | (0.00555) | (0.02938) | (0.00555) | (0.02932) | (0.00005) | (0.00026) |
| Province | Yes |  | Yes |  | Yes |  |
| Year | Yes | Yes | Yes | Yes | Yes | Yes |
| Observations | 133,122 | 133,122 | 133,122 | 133,122 | 133,122 | 133,122 |
| Adjusted $R^2$ | 0.4553 | 0.1113 | 0.4578 | 0.1152 | 0.8643 | 0.8630 |

This table presents the results of our tests analyzing the non-linear impact of access to credit on energy use, carbon emissions and energy efficiency of household. The square of Credit Access is introduced in regressions.
*Indicates significance at the 10% level.
**Indicates significance at the 5% level.
***Indicates significance at the 1% level.

Research shows that financial inclusion could lead to higher emission of carbon in the region (Le et al., 2020) and lower energy efficiency of manufacturing firms (Zhang et al., 2020). However, in the household sector, financial inclusion could improve the energy efficiency. Since households have insufficient voluntary efforts to reduce their carbon emissions in general (Dubois et al., 2019), it is essential to find external tools to meet carbon emission goals under the Paris Agreement. According to our results, broadening access to credit in financial market extends household choices in consumption categories, each of which is related to different energy and carbon intensity, and then has potential to spontaneously lead households to use energy with greater efficiency. The main policy implication of this paper is that policymakers need to encourage people to access credit market to better energy efficiency from household consumption. Although consumers might be indifferent to the energy implications of their purchases (Boardman, 2004), financial markets could find a way. Future work probing into the roles that household financial literacy (Korkmaz et al., 2021), digital finance (Yue et al., 2022), digital divide (Akhter, 2003) and demographic conditions (Lee et al., 2014) play in carbon emissions and energy efficiency would be an important contribution to the literature.

**CRediT authorship contribution statement**

**Jun Zhou:** Conceptualization, Writing – original draft, Writing – review & editing. **Zhichao Yin:** Supervision, Resources. **Pengpeng Yue:** Conceptualization, Methodology, Software, Formal analysis, Data curation, Writing – original draft, Visualization.

**Table A.1**
The calculated energy intensity in 2005–2019.

| Year | (1) Food | (2) Clothing | (3) Facilities | (4) Medicine | (5) Communication | (6) Education | (7) Residence | (8) Commodities |
|---|---|---|---|---|---|---|---|---|
| 2005 | 17.64 | 18.06 | 6.56 | 13.85 | 5.85 | 48.18 | 82.53 | 8.23 |
| 2006 | 14.82 | 16.59 | 5.57 | 12.41 | 5.15 | 43.93 | 69.95 | 7.23 |
| 2007 | 12.09 | 14.03 | 4.59 | 9.80 | 4.59 | 35.37 | 55.02 | 5.81 |
| 2008 | 11.22 | 12.60 | 4.66 | 9.16 | 5.18 | 33.61 | 57.58 | 5.56 |
| 2009 | 10.27 | 11.08 | 4.19 | 7.80 | 4.27 | 30.03 | 51.66 | 4.70 |
| 2010 | 9.63 | 10.83 | 3.99 | 7.32 | 4.34 | 28.06 | 48.52 | 4.11 |
| 2011 | 8.39 | 9.19 | 3.64 | 6.65 | 3.99 | 26.16 | 47.50 | 4.00 |
| 2012 | 11.69 | 10.22 | 3.78 | 8.70 | 1.99 | 27.65 | 51.17 | 5.24 |
| 2013 | 11.07 | 9.55 | 3.46 | 7.91 | 1.80 | 25.35 | 49.07 | 5.03 |
| 2014 | 9.48 | 8.36 | 3.18 | 7.29 | 1.51 | 22.07 | 48.64 | 4.01 |
| 2015 | 8.89 | 10.79 | 2.91 | 7.20 | 1.48 | 20.82 | 47.01 | 3.30 |
| 2016 | 8.66 | 10.21 | 2.64 | 6.44 | 1.50 | 19.91 | 44.89 | 2.81 |
| 2017 | 7.66 | 9.15 | 2.34 | 5.24 | 1.54 | 19.80 | 41.46 | 2.65 |
| 2018 | 7.06 | 6.40 | 2.20 | 4.33 | 2.15 | 18.49 | 38.17 | 2.48 |
| 2019 | 7.18 | 6.46 | 2.05 | 4.02 | 1.67 | 17.26 | 35.71 | 2.23 |

This table shows the calculated energy intensity ($GJ/10^4$ Yuan) of the related sectors in 2005–2019.

**Table A.2**
The calculated carbon intensity in 2005–2019.

| Year | (1) Food | (2) Clothing | (3) Facilities | (4) Medicine | (5) Communication | (6) Education | (7) Residence | (8) Commodities |
|---|---|---|---|---|---|---|---|---|
| 2005 | 163.68 | 167.91 | 59.45 | 128.84 | 52.34 | 451.22 | 753.62 | 76.43 |
| 2006 | 137.60 | 154.50 | 50.58 | 115.41 | 46.15 | 411.56 | 639.38 | 67.07 |
| 2007 | 112.34 | 130.70 | 41.72 | 90.88 | 41.18 | 331.35 | 501.36 | 53.89 |
| 2008 | 104.02 | 116.95 | 42.12 | 84.51 | 45.85 | 313.96 | 525.18 | 51.35 |
| 2009 | 95.06 | 102.95 | 37.73 | 71.69 | 38.10 | 280.93 | 469.50 | 43.37 |
| 2010 | 88.98 | 100.75 | 35.88 | 67.18 | 38.95 | 262.53 | 438.99 | 37.90 |
| 2011 | 77.62 | 85.69 | 33.03 | 61.05 | 35.77 | 245.25 | 429.18 | 36.75 |
| 2012 | 108.80 | 95.65 | 34.49 | 80.27 | 17.64 | 258.63 | 465.41 | 48.39 |
| 2013 | 102.84 | 89.38 | 31.52 | 72.85 | 15.92 | 236.46 | 445.02 | 46.37 |
| 2014 | 87.89 | 78.09 | 29.03 | 67.06 | 13.30 | 205.45 | 440.60 | 36.73 |
| 2015 | 82.37 | 100.99 | 26.60 | 66.22 | 13.07 | 193.07 | 421.86 | 30.03 |
| 2016 | 79.59 | 94.38 | 23.96 | 59.05 | 13.05 | 184.07 | 400.41 | 25.41 |
| 2017 | 69.62 | 83.24 | 21.00 | 47.70 | 13.29 | 180.87 | 366.31 | 23.65 |
| 2018 | 63.74 | 56.92 | 19.65 | 38.95 | 16.86 | 167.10 | 334.40 | 21.46 |
| 2019 | 63.92 | 56.59 | 18.41 | 35.69 | 14.20 | 155.50 | 311.67 | 19.05 |

This table shows the calculated carbon intensity ($kg/10^4$ Yuan) of the related sectors in 2005–2019.

**Declaration of competing interest**

The authors declare that they have no known competing financial interests or personal relationships that could have appeared to influence the work reported in this paper.

**Funding statement**

This work has been financially supported by the National Natural Science Foundation of China (72103010) and the National Social Science Fund of China (22CJY066).

**Data availability**

Data will be made available on request.

**Appendix. The calculated energy intensity and carbon intensity in 2005–2019**

See Tables A.1 and A.2.